# Modeling Infection with Multi-agent Dynamics

Wen Dong and Alex "Sandy" Pentland
M. I. T. Media Laboratory
M. I. T.
{wdong, sandy}@media.mit.edu

Katherine Heller
Department of Brain and Cognitive Sciences
M. I. T.
kheller@gmail.com

*Abstract* — Developing the ability to comprehensively study infections in small populations enables us to improve epidemic models and better advise individuals about potential risks to their health. We currently have a limited understanding of how infections spread within a small population because it has been difficult to closely track and infection within a complete community. This paper presents data closely tracking the spread of an infection centered on a student dormitory, collected by leveraging the residents' use of cellular phones. This data is based on daily symptom surveys taken over a period of four months and proximity tracking through cellular phones. We demonstrate that using a Bayesian, discrete-time multi-agent model of infection to model the real-world symptom report and proximity tracking records can give us important insights about infections in small populations.

## I. INTRODUCTION

Modeling contagions in social networks can help us facilitate the spread of valuable ideas and prevent disease. However, because closely tracking proximity and contagion in an entire community for months was previously impossible, modeling efforts have focused on large populations. As a result, we could say little about how an individual can better welcome good contagion and avoid bad contagion through his immediate social network. This paper describes how a "common" cold spread through a student residence hall community, with information based on daily surveys of symptoms for four months and tracking the locations and proximities of the students every six minutes through their cell phones. This paper also reports how infection occurred – and how infection could have been avoided – based on fitting the susceptible-infectious-susceptible (SIS) epidemic model to symptoms and proximity observations. It combines epidemic models and pervasive sensor data to give individually-tailored suggestions about local contagion, and also demonstrated the necessity of extending the epidemic model to individual-level interactions.

Epidemiologists agree on a framework for describing epidemic dynamics – people in a population can express different epidemic states, and change their states according to certain events. Computing event rates requires only knowledge about the overall-population at the present time. The susceptible-infectious-recovered (SIR) model, for example, divides the population into susceptible, infectious, and recovered sub-populations (or "compartments"). A susceptible person will be infected at a rate proportional to how likely the susceptible person is to make contact with an infected disease carrier, and an infected person will recover and gain lifetime immunity at a constant rate. Other compartmental models include the susceptible-infectious-susceptible (SIS) model for the common cold, in which infectious people become susceptible again once recovered, and the susceptible-exposed-infectious-recovered (SEIR) model, in which infected carriers experience an "exposed" period before they become infectious.

However, the availability of new data and computational power has driven model improvements, refining compartmental models that assume homogeneous compartments and temporal dynamics, to develop the Epidemiological Simulation System (EpiSimS) that take land use into account [1], and more recently simulations based on the tracking of face-to-face interactions in different communities [2][3][4][5][6].

These simulations all show evidence in favor of an epidemic dynamics framework, and against the assumption of homogeneous relationship and temporal dynamics. Using these kinds of algorithms with real-world symptom reports and proximity data could offer a much better understanding of how infection actually transfers from individual to individual, allowing for personalized contagion recommendations.

To understand the infection dynamics in a community at the individual level, we use the data collected in the Social Evolution experiment, part of which tracked "common cold" symptoms in a student residence hall from January 2009 to April 2009. The study monitored more than 80% of the residents of the undergraduate residence hall used in the Social Evolution experiment, through their cell phones from October 2008 to May 2009, taking daily surveys and tracking their locations, proximities and phone calls. This residence hall housed approximately 30 freshmen, 20 sophomores, 10 juniors, 10 seniors and 10 graduate student tutors. Researchers conducted monthly surveys on various social relationships, health-related issues, and status and political issues. They captured the locations and proximity of the students by instructing the cell phones to scan nearby Wi-Fi access points and Bluetooth devices every 6 minutes. They then collected the latitudes and longitudes of the Wi-Fi access points and the demographic data of the students to make sense of the data set. The data are protected by MIT COUIS and related laws.

This paper makes the following contributions to the field of human behavior modeling: It is among the first to discuss the spread of flu symptoms, tracked daily with cellphone-conducted surveys over an entire community. It is also among the first to model the spread of flu symptoms by looking at proximity tracked by cell phones, paired with a repository of other cellphone-conducted surveys about activity, status, and demographics. Lastly, this paper introduces a multi-agent model that is compatible with compartmental epidemic models and can infer who infected whom and how to avoid catching the flu. The large quantity of behavioral data generated from pervasive computing technology provides the details necessary to shift social sciences research from the level of large populations to individuals, and to enable social sciences to give more personalized advice.

The rest of the paper is organized as follows: In section II we describe the structure of face-to-face contact in the residence hall community, and the sensor data that captures this structure. In section III we introduce a Bayesian, multi-agent model, related to the Markov jump process, that not only simulates contagion but also makes inferences from observations. In section IV we demonstrate that we can effectively predict new cases of symptoms, identify cases of symptoms even if students do not report them, and determine the students and contacts that are most critical for symptom-spreading. Hence, we show that the multi-agent model captures how symptoms of the common cold and the flu spread in a student dormitory community.

## II. CONTAGION IN SOCIAL EVOLUTION EXPERIMENT

In this section, we discuss the structure of dynamic interpersonal interactions in the Social Evolution data, and the evidence it provides in support of applying an epidemic dynamics model at the individual level.

### A. Interactions in the residence hall community

Shared time, space, and previous relationships are the points from which the residents built new relationships, as shown by our monthly relationship surveys. Shared living sector in the dorm was the most important factor, especially for new residents. Shared courses and shared on-campus extra-curricular activities were also important factors.

The physical layout of the student dorm is the most important factor for the students in building relationships. We collected the room numbers of 71 out of the 84 residents, who live in eight living sectors separated by four floors and a firewall. A given student was five times more likely to report another student in his sector as a friend than another student living in a different sector. The 18 surveyed residents living in double rooms all reported their roommates as friends. At the beginning of a school year, a student on average socialized with half of the students living in his sector and 3% of the students living in different sectors. At the end of a school year, he socialized with about one third of students in the same sector and about 2% of the students in other sectors.

The students' academic curricula and extra-curricular activities were also important contexts through which they built relationships. A student reported another who was in the same year in school as a friend five times more often than a student in a different year. The average numbers of friends reported by freshmen, sophomores, juniors, seniors, and graduate tutors are respectively 2.9, 7.1, 8.1, 5.4, and 4.7 in the first month of school year. Students are less likely to report each other as friends if their years in school differ by more than one. Over time, every freshman made five friends on average, every graduate tutor made up to nine friends, and those who had already stayed in residence hall for more than a year (sophomores, juniors, and seniors) changed fewer than 10% of their friendship relations.

Figure 1 shows how the friendship network evolved from September 2008 to March 2009. Different living sectors are represented with different colors. An opening in the firewall connects dormitory sectors "f282.3" and "f290.3." Numbers 1 to 5 represents freshmen, sophomores, juniors, seniors, and graduate resident tutors, respectively. The relationships in September 2008 were based mostly on living sectors, because the freshmen met mostly those in the same sector, and the other residents had already adjusted their living sectors based on existing friendships. New relationships between September 2008 and March 2009 were mostly connected with the freshmen.

The monthly surveys also indicate that friends have a higher correlation in their on-campus activities. (the null hypothesis that friends and non-friends have the same probability distributions of empirical activity was rejected with $p < 10^{-6}$ in a Kolmogorov-Smirnov hypothesis test). From September 2008 to May 2009, 15 friend pairs each shared all on-campus activities that we surveyed, and 30% of friend pairs shared over 50% of their on-campus activities. In comparison, non-friends shared less than 10% of on-campus activities.

### B. Spreading of symptoms in the residence hall community

In the Social Evolution experiment, we offered students $1 per day from 01/08/2009 through 04/25/2009 to answer surveys about contracting the flu, regarding the following specific symptoms: (1) runny nose, nasal congestion, and sneezing; (2) nausea, vomiting, and diarrhea; (3) frequent stress; (4) sadness and depression; and (5) fever. Altogether, 65 residents out of 84 answered the flu surveys, each of whom

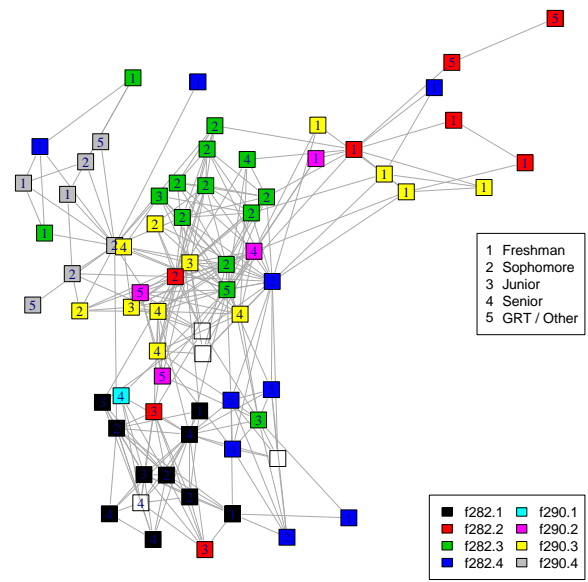

**Figure 1: Subjects in the dormitory formed clusters of relationships by their dormitory sectors (the primary factor) and their years in school (the second most important factor).**

answered for half of the surveyed period. The correlation between stress and sadness is 0.39, while the correlations between other pairs are about 0.10.

The symptom self-reporting in the Social Evolution data seems to be compatible with what the epidemic model would indicate: symptoms other than runny nose are probabilistically dependent on that student's friendship network. The durations of symptoms were about two days, and fit the exponential distribution well. The chance of reporting a symptom is about 0.01, and each individual had a 0.006~0.035 increased chance of reporting a symptom for each additional friend with the same symptom. These parameters are useful for epidemic simulation in the residence hall network, and for setting the initial values of fitting an epidemic model to real-world symptom observations and sensor data. The symptom surveys show some repeated infections, several clustered infections, the persistence of infections in larger clusters, and the persistence of infections caused by individuals who took longer to recover.

In this data, a student with a symptom had 3-10 times higher odds of seeing his friends with the same symptom (again, except for runny nose). As such, it makes sense to fit the time-tested infection model with real-world data of symptom reports and proximity observations, and infer how friends infect one another through their contacts. In order to determine whether the higher odds could somehow be due to chance, we conducted the following permutation test to reject the null hypothesis that "the friendship network is unrelated to symptoms," and we can reject that null hypothesis with $p < 0.05$. The permutation test shuffles the mapping between the students and the nodes in the friendship network and estimates the probability distribution of the number of friends with the same symptom among all possible shuffling. If friendship networks were not related to the timing of when a student exhibits a symptom, then all mappings between the students and the nodes would be equally likely, and the number of friends with the same symptom would take the more likely values.

Because the epidemic model says that an infected/infectious individual recovers with a constant rate γ, and therefore the duration of the infection follows an exponential distribution, we fit the observed symptom durations with exponential distribution using maximum likelihood estimation to check the compatibility of the epidemic model with the Social Evolution data on recovery rate. The average durations of the symptoms in the fit exponential distributions were 2.9 days for runny nose, nasal congestion, and sneezing, 2.2 days for stress, 1.7 days for sadness and depression, 1.4 days for nausea, vomiting, and diarrhea, and 1.4 days for fever. As such, we have satisfactory goodness-of-fit statistics ($p \approx 0.6$ in Kolmogorov hypothesis testing).

Further, the probability of exhibiting a symptom increased approximately linearly with the number of friends reporting the same symptom. The base probability of having a symptom was 1~2 cases per day per 100 persons, and every additional friend exhibiting the symptom added the probability by about 1%. This relationship again agrees with the theory of epidemic dynamics, which predict that the rate of contagion will be proportional to the likelihood of contact with an infected individual (Table 1).

**Table 1: Probability of catching symptom = $1 - \alpha \times exp(-\beta \times$ number of neighbors with symptom), $R^2$ and $p$.**

| Symptom | $\alpha$ | $\beta$ | $R^2$ | $p$ |
|---|---|---|---|---|
| Runny nose | 1.013 | 0.024 | 0.52 | 0.04 |
| Sadness | 0.991 | 0.016 | 0.63 | 0.13 |
| Stress | 1.001 | 0.035 | 0.85 | 0.005 |
| Nausea | 0.993 | 0.006 | 0.94 | 0.11 |

III. MODELING INFECTION DYNAMICS

In this section, we take a discrete-time stochastic multi-agent SIS model (Algorithm 1), and derive an inference algorithm to fit this multi-agent model to real-world data on proximity and symptom reporting. The inference algorithm does three things. First, it learns the parameters of the multi-agent model, such as rate of infection and rate of recovery. Second, it estimates the likelihood that an individual was infectious from the contact he had with other students, and from whether those others reported symptoms when the individual's symptom report is not available. Finally, it enables us to make useful predictions about contracting infections within the community in general. We then discuss how this multi-agent model (which works at the level of individual participants) is related to the traditional compartmental model (which works at aggregate level): what this multi-agent model gives us, how the discrete-time version of the multi-agent model gives us a much simpler inference algorithm in comparison to its continuous-time counterpart, and when the discrete-time model approximates the continuous-time counterpart poorly.

**Algorithm 1: Discrete-time stochastic multi-agent SIS model to fit real-world infection dynamics.**

- Input:

  o $G = (N, E)$ : dynamic network where nodes $N = \{1, ..., number\ of\ people\}$ representing people, bi-directional edges $E = \{(n_1, n_2, t): n_1\ is\ near\ n_2\ at\ time\ t\}$ representing "nearby" relation, and $t \in \{1, ..., T\}$.

  o Prior knowledge that $\alpha$ — probability that an infectious person outside of the network makes a susceptible person within the network infectious, $\beta$ — probability that an infectious person within the network makes a susceptible nearby person infectious, and $\gamma$ — probability that an infectious person becomes susceptible, all take their likely values according to beta distributions with known hyper parameters. $\alpha \sim beta(a, b), \beta \sim beta(a', b'), \gamma \sim beta(a'', b'')$

  o Prior knowledge on which symptoms are dependent on common cold state $P(Y|X)$.

- Output: a matrix structure $\{X_{n,t}, Y_{n,t}: n, t\}$ indexed by time $t$ and node $n$. The state $X_{n,t}$ of node $n$ at time $t$ is either 0 (susceptible) or 1 (infected). The symptom $Y_{n,t}$ of node $n$ at time $t$ is probabilistically dependent on the state of node $n$ at time $t$. A collection of events $R$ drives change in state.

- Procedure:
  - Sample parameters according to beta distributions. $\alpha \sim beta(a,b), \beta \sim beta(a',b'), \gamma \sim beta(a'',b'')$.
  - $X_{n,1} = 0, \forall n \in N$: all people are susceptible at $t = 1$.
  - At each time $t + 1 = 2, \ldots, T$:
    - Infectious person becomes susceptible with probability $\gamma$: $\{X_{n,t} = 1 \to X_{n,t+1} = 0\} \sim \text{Bernoulli}(\gamma)$. If the Bernoulli trial is a success, $R \leftarrow R \cup \{X_{n,t} = 1 \to X_{n,t+1} = 0\}$, set $X_{n,t+1} \leftarrow 0$, and set $Y_{n,t+1} \sim P(Y|X = X_{n,t+1})$ accordingly.
    - Infectious persons within and outside of the network contribute to turning a susceptible person infectious, and the contributions happen independently:
    
    $$\{X_{n,t} = 0, X_{n',t} = 1, (n',n,t) \in E \to X_{n,t+1} = 1\} \sim \text{Bernoulli}(\beta)$$
    $$\{X_{n,t} = 0 \to X_{n,t+1} = 1\} \sim \text{Bernoulli}(\alpha)$$
    
    Set $X_{n,t+1} \leftarrow 1$ if any of the above Bernoulli trials is a success, and set $Y_{n,t+1} \sim P(Y|X = X_{n,t+1})$ accordingly. Add the successful trials into $R$.

Algorithm 1 defines a generative discrete-time stochastic multi-agent SIS model of infection dynamics. It is an **SIS** model because each susceptible individual can turn infectious through a close contact with an infectious individual, and each infectious individual can recover and again be susceptible. It is **generative** in the sense that we can use this model to generate a time series of the susceptible/infectious states of individuals given a dynamic network and set of parameters. It is **stochastic** because each time series is associated with a probability defined by how likely it is that this generative process will generate this time series. It is a **multi-agent** model, and copes with each individual differently according to his instantaneous connections and the instantaneous susceptible/infectious states of the individuals currently connected with him. In contrast, a compartmental model works at the population level, and treats individuals with the same states similarly. We use a **discrete-time approximation** of a Markov jump process, because our data is discretized.

We denote the (state-changing) events using a production system, and introduce minimum notation to differentiate individuals. In place of $I \to S$ (an infectious individual is recovered and becomes susceptible), we write $\{X_{n,t} = 1 \to X_{n+1,t} = 0\}$ (Individual $n$ is infectious at time t and becomes susceptible at time $t + 1$). In place of $S + I \to 2 \times I$, we write $\{X_{n,t} = 0, X_{n',t} = 1, (n,n',t) \in E \to X_{n,t+1} = 1\}$ (Individual $n'$ contributes to the infection of individual $n$ through their contact at time $t$). In place of $S \to I$, we write $\{X_{n,t} = 0 \to X_{n+1,t} = 1\}$ (Individual is infected from outside of the network).

We derive the probability of a sample path generated by our multi-agent model, such that we can proceed to make inferences with this model. The probability that a person is infected between $t$ and $t + 1$ is the (marginal) probability that either (1) any of his infectious connections contributed to infect him, or (2) infectious people outside of the network infected him:

$$P(X_{n,t+1} = 1 | X_{n,t} = 0, \{X_{n',t}: n' \neq n\})$$
$$= 1 - P(X_{n,t+1} = 0 | X_{n,t} = 0, \{X_{n',t}: n' \neq n\})$$
$$= 1$$
$$- P\left(\cap_{n'} \overline{\{X_{n,t} = 0, X_{n',t} = 1, (n,n',t) \in E \to X_{n+1,t} = 1\}} \cap \overline{\{X_{n,t} = 0 \to X_{n+1,t} = 1\}}\right)$$
$$= 1$$
$$- \prod_{n'} P\left(\overline{\{X_{n,t} = 0, X_{n',t} = 1, (n,n',t) \in E \to X_{n+1,t} = 1\}}\right)$$
$$\cdot P\left(\overline{\{X_{n,t} = 0 \to X_{n+1,t} = 1\}}\right), \text{ independence of events}$$
$$= 1 - (1-\alpha)(1-\beta)^{\sum_{n':(n',n,t) \in E} X_{n',t}}, \text{ by SIS model}$$
$$\approx \alpha + \beta \cdot \sum_{(n',n,t) \in E} X_{n',t}, \text{ when } \alpha, \beta \cdot \sum_{(n',n,t) \in E} X_{n',t} \ll 1$$

In the above, over-line represents set complement. When the probability of infection is small, it is approximately the sum of the probabilities from different sources (related to $\alpha$ and $\beta$), and the probability that more than one source contributed to infection is small. This approximation works well for our data set.

The probability that a collection of events $R \subseteq \{X_{n,t} = 0 \to X_{n+1,t} = 1\} \cup \{\{X_{n,t} = 0, X_{n',t} = 1, (n,n',t) \in E \to X_{n+1,t} = 1\}: n'\}$ makes a person infected is the product of the probabilities of the events in $R$, because the events are conditionally independent given the current state, normalized.

$$P(R|X_{n,t} = 0, X_{n,t+1} = 1)$$
$$= \frac{P(R|X_{n,t} = 0)}{P(X_{n,t+1} = 1 | X_{n,t} = 0)}, \text{ because } P(X_{n,t+1} = 1 | X_{n,t} = 0, R) = 1$$
$$= \frac{\prod_{r \in R} P(r|X_{n,t} = 0)}{P(X_{n,t+1} = 1 | X_{n,t} = 0)}, \text{ independence of events}$$
$$= \alpha^{1_{\{X_{n,t}=0 \to X_{n+1,t}=1\} \in R}} \cdot \prod_{n' \neq n} \beta^{1_{\{X_{n,t}=0, X_{n',t}=1, (n,n',t) \in E \to X_{n+1,t}=1\} \in R}}$$
$$\cdot \alpha^{1_{\{X_{n,t}=0 \to X_{n+1,t}=1\} \notin R}} \cdot \prod_{n' \neq n} (1-\beta)^{1_{\{X_{n,t}=0, X_{n',t}=1, (n,n',t) \in E \to X_{n+1,t}=1\} \notin R}}$$
$$\cdot \frac{1}{P(X_{n,t+1} = 1 | X_{n,t} = 0)}$$

In the above, the function 1 is a characteristic function. It is 1 when its condition is true, 0 otherwise.

The state-event correspondences in the other cases are determined by the lack of events: keeping the infectious/susceptible state between two consecutive time steps corresponds to no infection/recovery event, and recovery corresponds to a unique recovery event.

The probability of seeing a state sequence/matrix $\{X_{n,t}: n,t\}$ is therefore

$$P(\{X_{n,t}: n,t\}, \alpha, \beta, \gamma)$$

$$= P(\alpha)P(\beta)P(\gamma) \prod_n P(X_{n,1}) \prod_{t,n} P(X_{n,t+1}|(X_{n',t}),\alpha,\beta,\gamma)$$

$$= P(\alpha)P(\beta)P(\gamma) \prod_{t,n} \gamma^{1_{X_{n,t}=1} \cdot 1_{X_{n,t+1}=0}} \cdot (1-\gamma)^{1_{X_{n,t}=1} \cdot 1_{X_{n,t+1}=0}}$$

$$\cdot \left(\alpha + \beta \cdot \sum_{(n',n,t) \in E} X_{n',t}\right)^{1_{X_{n,t}=0} \cdot 1_{X_{n,t+1}=1}}$$

$$\cdot \left(1 - \alpha - \beta \cdot \sum_{(n',n,t) \in E} X_{n',t}\right)^{1_{X_{n,t}=0} \cdot 1_{X_{n,t+1}=0}}$$

The second step above is due to the Markov property and to the fact that the state changes of the individuals are independent Bernoulli trials, both induced by the generative SIS process. In the third step above, the four factors behind the product sign are shorthand that expresses the probabilities by four different cases. 1 is a characteristic function.

The probability of seeing a collection of events $R$ is the probability of the state sequence given by $R$ multiplied by the conditional probability that $R$, instead of another collection, generated the state sequence. The probability of seeing $R$ can also be expressed as the product of the probabilities of infection/recovery events in $R$ times the product of the probabilities that individuals didn't change their states, normalized. The events in $R$ tell us the critical interactions in the network. We skip the formula for simplicity.

We incorporate symptom observation in the multi-agent SIS model in order to make inferences such as whether an observed symptom is due to infection according to our multi-agent SIS model or to something else, what implications about epidemics can be drawn from incomplete symptom reports, and how different interventions could help. We specify that the observations $\{Y_{n,t}: (n,t) \in obs \subseteq N \times \{1,...,T\}\}$ are mutually independent given the corresponding latent susceptible/infectious states. $P(\{X_{n,t}, Y_{n,t}: n, t\}, \alpha, \beta, \gamma) = P(\alpha)P(\beta)P(\gamma) \cdot \prod_n P(X_{n,1}) \cdot \prod_{t,n} P(X_{n,t+1}|(X_{n',t}), \alpha, \beta, \gamma) \cdot \prod_{(t,n) \in obs} P(Y_{n,t}|X_{n,t})$. We do not expand the formula, for simplicity.

We employ a Gibbs sampler to iteratively sample infectious/susceptible state sequence from Bernoulli distributions, sample events conditioned on state sequence from Bernoulli distributions, and sample parameters from Beta distributions (Algorithm 2).

A Gibbs sampler is an algorithm to generate a sequence of samples from the joint probability distribution of two or more random variables. The Gibbs sampling algorithm generates an instance from the distribution of each variable in turn, conditional on the current values of the other variables. It can be shown that the sequence of samples constitutes a Markov chain, and that the stationary distribution of the Markov chain is the sought-after joint distribution only.

**Algorithm 2: Gibbs sampler**

- Input:

  o $G = (N, E)$ : dynamic network where node $N = \{1, ..., number\ of\ people\}$, bi-directional edges $E = \{(n_1, n_2, t): n_1\ is\ near\ n_2\ at\ time\ t\}$ representing "nearby" relation, and $t \in \{1, ..., T\}$.

  o $\{Y_{n,t}: (n,t) \in obs \subseteq N \times \{1,...,T\}\}$: observation matrix of symptoms indexed by time and node.

- Output: a sample of the parameters and the state matrix of the stochastic infection process.

  o Parameters: base rate of infection ($\alpha$), rate of infection by each additional infectious neighbor ($\beta$), rate of recovery ($\gamma$), emission matrix $P(Y_{n,t}|X_{n,t})$ (probability of emitting a specific observation under a specific infectious/susceptible latent state).

  o $\{X_{n,t}: n, t\}$: state matrix indexed by time and node. The state $X_{n,t}$ of node $n$ at time $t$ is either 0 (susceptible) or 1 (infected).

- Procedure: initialize parameters and state matrix (section II-B), then alternate between sampling latent common-cold state, sampling infection/recovery events and sampling parameters until convergence.

$$X_{n,t+1}|\{X_{n,t}: n, t\} \backslash X_{n,t+1}; \alpha, \beta, \gamma$$

$$\sim Bernoulli\left(\frac{P(X_{n,t+1}=1, \{X_{n,t}, Y_{n,t}: n, t\} \backslash X_{n,t+1})}{P(X_{n,t+1}=0, \{X_{n,t}, Y_{n,t}: n, t\} \backslash X_{n,t+1})}\right)$$
$$+P(X_{n,t+1}=1, \{X_{n,t}, Y_{n,t}: n, t\} \backslash X_{n,t+1})$$

$$\{\circ\} \cup \{n': (n, n', t) \in E\} | X_{n,t} = 0, X_{n,t+1} = 1$$

$$\sim Categorical\left(\frac{\alpha, \beta, ..., \beta}{\alpha + \beta \sum_{n'} 1_{(n',n,t) \in E \cap X_{n',t}=1}}\right)$$

We use $\circ$ and $n'$ to represent different infection sources.

$$\{X_{n,t} = 1 \rightarrow X_{n,t+1} = 0\} | X_{n,t} = 1, X_{n,t+1} = 0$$
$$\sim Bernoulli(1)$$

$$\alpha \sim Beta\left(a + \sum_{n,t} 1_{\{X_{n,t}=0 \rightarrow X_{n,t+1}=1\} \in R}, b + \sum_{n,t} 1_{X_{n,t}=0}\right.$$
$$\left. - \sum_{n,t} 1_{\{X_{n,t}=0 \rightarrow X_{n,t+1}=1\} \in R}\right)$$

$$\beta \sim Beta\left(a' + \sum_{n,t} 1_{\{X_{n,t}=0, X_{n',t}=1, (n',n,t) \in E \rightarrow X_{n,t+1}=1\} \in R}, b'\right.$$
$$+ \sum_{n,t} 1_{(n',n,t) \in E \cap X_{n',t}=1 \cap X_{n,t+1}=1}$$
$$\left. - \sum_{n,t} 1_{\{X_{n,t}=0, X_{n',t}=1, (n',n,t) \in E \rightarrow X_{n,t+1}=1\} \in R}\right)$$

$$\gamma \sim Beta\left(a'' + \sum_{n,t} 1_{\{X_{n,t}=1 \rightarrow X_{n,t+1}=0\} \in R}, b'' + \sum_{n,t} 1_{X_{n,t}=1}\right.$$
$$\left. - \sum_{n,t} 1_{\{X_{n,t}=1 \rightarrow X_{n,t+1}=0\} \in R}\right)$$

The SIS model describes infection dynamics in which the infection doesn't confer long-lasting immunity, and so an individual becomes susceptible again once recovered. The common cold has this infection characteristic.

This discrete-time multi-agent model of susceptible-infectious-susceptible (SIS) dynamics specializes in paths of infection and individual-level interaction. It is a very appropriate individual-level model for our discrete data, where the underlying infection/recovery times are exponentially distributed (as discussed in section II). In contrast, the differential equation model and the stochastic equations of SIS dynamics work at the population level, and their variables are respectively densities and sizes of susceptible and infectious populations. The differential equation model

$$\dot{S} = -\beta \cdot SI + \gamma \cdot I$$
$$\dot{I} = \beta \cdot SI - \gamma \cdot I$$

for SIS specifies that the rate of change of the infectious-population density is bilinear for both the infectious-population density and susceptible-population density. In this system, two individuals from the infectious population are the same, two individuals from the susceptible population are the same, and two individuals from different populations have an equal chance of causing infection. The stochastic model

$$S + I \rightarrow 2I, rate = \beta' \cdot |S| \cdot |I|$$
$$I \rightarrow S, rate = \gamma' \cdot |I|$$

specifies that infection happens at a rate that is bilinear in both the infectious-population density and the susceptible-population density. The stochastic model enables us to reason about the randomness in the SIS system when the population size is small and randomness cannot be ignored.

We are more interested in explaining the symptom observations in a community with susceptible-infectious-susceptible dynamics at a point in time. How likely was a person to be infectious at time t, given the number of his friends reporting symptoms, reporting no symptoms, and not answering surveys, on and after time t, given the infectious person's survey answers or that he didn't answer surveys, and given his recent proximity with his friends? How likely is a person to be infected? Which nodes and links were critical in spreading infection in the community? How do we control infection in this community?

The discrete-time multi-agent model approximates its continuous-time multi-agent counterpart well only when the time step size in the discrete-time model is smaller than the average rate of the infection/recovery events. The continuous-time multi-agent model is a Markov jump process, also known as a compound Poisson process. In a Markov jump process, the state $\{S(t): t\}$ changes according to a set of reactions $\{R_1, \dots, R_n\}$ that happen with rates dependent on current state $r_1(S(t)), \dots, r_n(S(t))$. The probability of a given event sequence $R_{j_1}, \dots, R_{j_m}$ happening at time $0 < t_1 < \cdots < t_m < T$ is $\prod_i r_{j_i}(S(t_i)) \exp(-\sum_{i=1}^{m}\sum_{j=1}^{n} r_j(S(t_i)) \cdot (t_{i+1} - t_i))$, where the product of event rates contains the "information" at discrete times when reactions happen, and where the exponential function contains the "information" when no reaction happens. Since the best discrete-time approximation of the continuous-time probability has to round the event times to the closest sample time, the approximation may be unsatisfactory when $S(t)$ changes quickly relative to the sample interval.

IV. EXPERIMENTAL RESULT

In this section we model the contagion which existed in the residence hall community. We estimate, at the community level, the parameters of susceptible-infectious-susceptible (SIS) infection dynamics – the probability with which an infectious individual recovers, the probability that a susceptible individual becomes infectious through an infection outside of the residence hall community, and the probability that a susceptible individual becomes infectious per day per infectious friend. At the individual level, we describe the results of using the Gibbs sampling algorithm to fit the discrete-time multi-agent SIS infection dynamics to symptom observations.

We describe the performance of our multi-agent model in predicting missing data in synthetic time series with different rates of infection and recovery and different levels observation noises, as well as how our model compares with a general-purpose algorithm such as support vector classifier. Then, we show that we can make useful inferences by fitting our multi-agent model with the hourly proximity records and the self-reported symptoms in the Social Evolution data set. Experimenting on synthetic data enables us to see how we can improve the performance of algorithms in interpreting epidemic time series, and what we expect them to be able to model well. While we do not have a daily clinical diagnosis of the flu and common cold, we can look at the statistics after we fill in the missing data by fitting the multi-agent model to proximity data and symptom self-reporting and see if they reflect any unexpected events, or agree with common knowledge about the cold found elsewhere.

*A. Calibrating performance*

We took several steps to calibrate the performances of the multi-agent model and support vector classifier on synthetic data. First, we synthesized 50 time series – each 128 days long – from the Bluetooth proximity pattern in the Social Evolution data and different parameterizationss. Then, we randomly removed the infectious/susceptible data from 10% of the population, added noise to the remaining data in each time series, and averaged the performances on inferring the held-out data corresponding to each method and parameterization.

The different sets of parameterization are (1) $\alpha = 0.01, \beta = 0.02, \gamma = 0.3$ and the observation error being 0.01; (2) $\alpha = 0.01, \beta = 0.02, \gamma = 0.3$ and the observation error being 0.001; (3) $\alpha = 0.005, \beta = 0.035, \gamma = 0.3$ and the observation error being 0.01. Comparing performances between (1) and (2) enables us to see the effect of observation error on algorithm performance, comparing performances between (1) and (3) enables us to see the effect of the network on algorithm performance, and comparing performances within each case enables us to see the difference between model-based learning and the black-box classifier.

We ran Gibbs samplers for 10,000 iterations, got rid of the initial 1000 burn-in iterations, and treated the remaining 9000 iterations as samples from the posterior distribution. We trained the support vector classifier from another 1000-day time series synthesized using the right parameterization, and used the number of infectious contacts yesterday, today, and tomorrow as a feature. We assigned different weights to the "infected" class and the "susceptible" class to balance the true prediction rate and the false prediction rate.

All methods can easily identify 20% of infectious cases in the missing data with little error, but the model-based method using our dynamic multi-agent system consistently performs better than the support vector classifier. Less noise in symptom observation and in the individuals' contact networks significantly improves the performance of inferring missing data, as shown through the ROC curves in Figure 2. An ROC (receiver operating characteristic) curve shows how different algorithms trade off between the true positive rate (the number of correctly-predicted positive cases divided by the total number of positive cases) and the false positive rate (the number of incorrectly-predicted positive cases over the total number of negative cases) in inferring the infectious cases for the 10% of the population whose states are missing. The curve indicates better performance if it correctly predicts more positive cases and incorrectly predicts fewer negative cases, or equivalently if it is closer to the top-left corner, or it has the larger area below.

The support vector classifier performs worse – especially in identifying the isolated infectious cases in the missing data – because it assumes that its cases are i.i.d (identical and independently distributed) and because including the temporal structure of epidemic dynamics into the features is not an easy task. The support vector classifier also assumes that we either already have enough training data or can synthesize training data. This assumption generally cannot be satisfied for the kinds of problems we are interested in here.

Observation noise not only makes inferring the individual states (whether a person is susceptible or infectious) difficult, but also increases uncertainty in the parameters of the whole system, which in turn also makes inferring individual states more difficult. If many susceptible cases were wrongly inferred as infectious, the estimated infection rate ($\alpha$ and $\beta$) of the whole system would be higher than it should be, and more susceptible cases would be wrongly inferred as infectious. Similarly, if many infectious cases were wrongly inferred as susceptible, the estimated recovery rate ($\gamma$) of the system would be higher than it should be, and one stretch of infectious cases would be split into many stretches. As such, observation noise reduces the model's ability to distinguish between susceptible states and infectious states.

Knowledge of the dynamical contact network also affects the quality of parameter estimation and (susceptible versus infections) state inference. The more we know about who contacted whom, the more targeted we can be in locating the infectious cases.

*B. Inferring common cold from symptom report and proximity*

In this section, we report several interesting statistics derived from the dynamical proximity network and symptom report and from the assumption of susceptible-infectious-susceptible dynamics. In order to infer latent common cold time series that best fits the multi-agent SIS model from dynamical Bluetooth proximity information and symptom self-report in the Social Evolution data using our Gibbs sampler (Algorithm 2), we extracted the hour-by-hour proximity snapshot over the 107 days we were monitoring symptoms and interpolated the hourly symptom report as the submitted daily symptom report. Due to the lack of prior knowledge on the joint probability distribution of different symptoms, we assumed that the symptoms are probabilistically independent given the common cold state. We ran the Gibbs sampler for 10,000 iterations, removed the first 1000 burn-in iterations, and took the rest as samples of the posterior probability distribution of common cold states conditioned on symptom self-reports. To shorten the burn-in process, we initialized the Gibbs sampler with $\alpha = 0.02, \beta = 0.02, \gamma = 0.3$, observation error being 0.01, and a common-cold state being "infectious" if more than two symptoms were reported. Initialization shouldn't matter if we can run the Gibbs sampler until convergence, because theoretically a Gibbs sampler will have attained the equilibrium state.

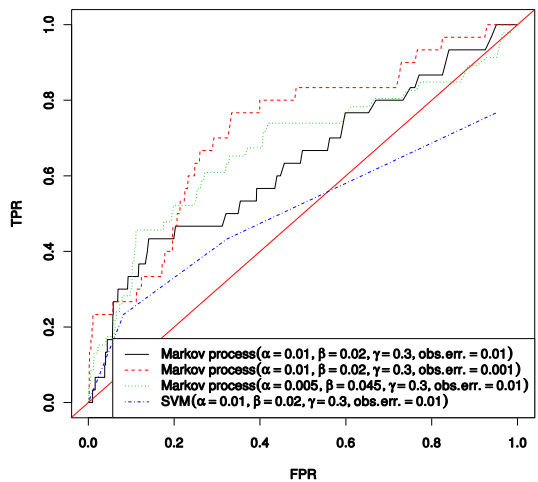

**Figure 2: Less observation error (obs.err.=0.001) and better knowledge about network ($\beta = 0.045$) lead to better trade-off between true positive rate (TPR) and false positive rate (FPR). The support vector classifier has worse trade-off between TPR and FPR than the multi-agent Markov model.**

We do not have the clinical truth of common cold diagnoses. However, the statistics that we discuss below make us believe that the multi-agent SIS model captures the structure of a diffusion process accompanying the symptom report.

Figure 3 shows the (marginal) likelihood of the daily common-cold states of individuals. Rows in this heat map are indexed by subjects, arranged so that friends go together, and are placed side by side with a dendrogram that organizes friends hierarchically into groups according to the distance between the individuals and groups. Different colors on the leaves of the dendrogram represent different living sectors in the student dorm. Columns in this heat map are indexed by date in 2009. Brightness of a heat-map entry indicates the likelihood of being infectious. The brighter a cell is, the more likely it is that the corresponding subject is infectious on the corresponding day. Sizes of black dots represent the number of reported symptoms, ranging from zero symptoms to all symptoms. When a black dot doesn't exist on the corresponding table entry, the corresponding person didn't answer the survey on the corresponding day.

This heat map shows clusters of common cold happenings, and in each cluster a few individuals reported symptom. When interpersonal proximities happened in larger social clusters, symptom clusters lasted longer and involved more people. A study of the heat map also tells us what the Gibbs sampler does in fitting the multi-agent SIS model to the symptom report: a subject often submitted flu-symptom surveys daily when he was in a "susceptible" state, but would forget to submit surveys when he was in the "infectious" state. The Gibbs sampler will nonetheless say that he was infectious for these days, because he was in the infectious state before and after, an infectious state normally lasts four days, and many of his contacts were in the infectious state as well. A subject sometimes reported symptoms when none of his friends did in the time frame. The Gibbs sampler will say the he was in the susceptible state, because the duration of the symptom reports didn't agree with the typical duration of a common cold, and because his symptom report was isolated in his contact network.

The inferred infectious state from symptom reports and hourly proximity networks normally lasts four days, but could be as long as two weeks. A student often caught a cold 2 ~ 3 times from the beginning of January to the end of April. The bi-weekly searches of the keyword "flu" from January 2009 to April 2009 in Boston – as reported by Google Trends – explains 30% of variance in the number of (aggregated) bi-weekly common cold cases inferred by the Gibbs sampler, and network size explains another 10%.

The timing of different symptoms with regard to the inferred common cold cases follows interesting patterns. Stress and sadness normally began three days before the onset of a stretch of infectious state, and lasted two weeks. Runny nose and coughing began zero to two days before the onset of a symptom report and ended in about seven days, and they have similar density distributions. Fever normally occurred on the second day after the onset of a stretch of infectious state, and lasted for about two days. Nausea often happened four days before the onset of reaching an infectious state, then disappeared and reappeared again at the onset.

## V. CONCLUSIONS

The study of infection in a small population has important implications both for refining epidemic models and for advising individuals about their health. The spread of infection in this context is poorly understood because of the difficulty in closely tracking infection in a complete community. This paper showcases the spread of an infection centered on a student dormitory, based on daily symptom surveys over a period of four months and on proximity tracking through resident cellular phones. It also demonstrates that fitting a discrete-time multi-agent model of infection with real-world symptom self-reports and proximity observations gives us useful insight in infection paths and infection prevention.

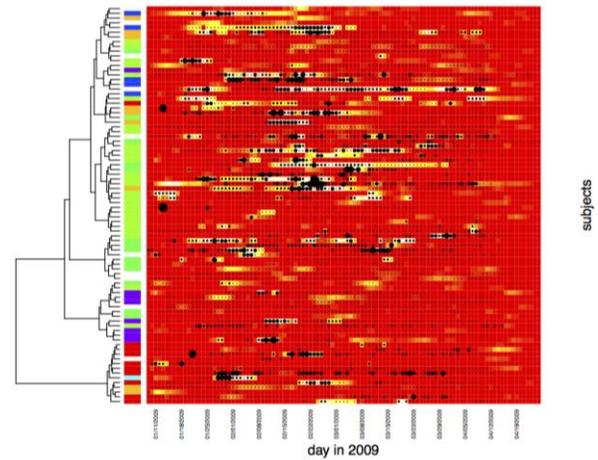

Figure 3: An agent-based model can infer common cold state, and captures infection from symptom self-report and proximity network. Sizes of black dots represent the number of symptoms reported, ranging from zero symptoms to all symptoms, and no black dot means no self-report.

ACKNOWLEDGMENT

Research was sponsored by the Army Research Laboratory under Cooperative Agreement Number W911NF-09-2-0053, and by AFOSR under Award Number FA9550-10-1-0122. Views and conclusions in this document are those of the authors and should not be interpreted as representing the official policies, either expressed or implied, of the Army Research Laboratory or the U.S. Government. The U.S. Government is authorized to reproduce and distribute reprints for Government purposes notwithstanding any copyright notation. Katherine Heller was supported on an NSF postdoctoral fellowship.